\documentclass{svproc}
\usepackage{graphicx}%
\usepackage{multirow}%
\usepackage{amsmath,amssymb,amsfonts}%
\usepackage{mathrsfs}%
\usepackage[title]{appendix}%
\usepackage{xcolor}%
\usepackage{textcomp}%
\usepackage{manyfoot}%
\usepackage{booktabs}%
\usepackage{algorithm}%
\usepackage{algorithmicx}%
\usepackage{algpseudocode}%
\usepackage{listings}%
\usepackage[colorinlistoftodos]{todonotes}
\usepackage[normalem]{ulem}
\usepackage{url}

\def\orcidID#1{\unskip$^{[#1]}$} 

\begin{document}
\mainmatter              
\title{Synthetic Networks That Preserve Edge Connectivity}
%
%
\author{Lahari Anne\inst{1}\orcidID{0009-0007-5708-2405} \and
The-Anh Vu-Le\inst{1}\orcidID{0000-0002-4480-5535} \and
Minhyuk Park\inst{1}\orcidID{0000-0002-8676-7565} \and
Tandy Warnow\inst{1}\orcidID{0000-0001-7717-3514}
\and
George Chacko\inst{1}\orcidID{0000-0002-2127-1892}
}
\authorrunning{L. Anne et al.}
\titlerunning{Improving SBM network generation}
%

\institute{Siebel School of Computing and Data Science, Grainger College of Engineering, University of Illinois Urbana-Champaign, IL 61801, USA\\
\email{correspondence: \{warnow \hspace{-1.5mm}| \hspace{-3mm} chackoge\}@cs.illinois.edu}}
\maketitle
\begin{abstract}
Since true communities within real-world networks are rarely known, synthetic networks with planted ground truths are valuable for evaluating the performance of community detection methods. 
Of the synthetic network generation tools available, Stochastic Block Models (SBMs) produce  networks with ground truth clusters that well approximate input parameters from real-world networks and clusterings. 
However, we show that SBMs can produce disconnected  ground truth clusters, even when given parameters from clusterings where all clusters are connected.   Here we describe the REalistic Cluster Connectivity Simulator (RECCS), a technique that modifies an SBM synthetic network to improve the fit to a given clustered real-world network with respect to  edge connectivity within clusters, while maintaining the good fit with respect to other network and cluster statistics.  Using  real-world networks up to 13.9 million nodes in size, we show  that RECCS, applied to stochastic block models, results in synthetic networks that have a better fit to cluster edge connectivity than unmodified SBMs, while providing roughly the same quality fit for other network and clustering parameters as  unmodified SBMs.  

\end{abstract}


\section{Introduction}\label{sec:Intro}

Community detection methods resolve networks at the meso-scale by identifying clusters of nodes that exhibit desirable properties, such as density, edge connectivity, and separability from the remainder of the network  \cite{Fortunato2022,ElMoussaoui2019,Javed2018,Coscia2011,Traag2019,Park2024}.   
Community detection (clustering) is used in many applications, therefore, an understanding of when a method is likely to produce clusters of good quality is of some importance.

Since ground truth communities are not reliably known in real-world networks, evaluation using synthetically generated networks with planted ground truth communities provides a useful alternative \cite{Peel2017}. 
Several synthetic network generators are in  use,  including Stochastic Block Models (SBM)\cite{peixoto_graph-tool_2014,Peixoto-chapter-2019},   the LFR  (Lancichinetti-Fortunato-Radicchi) generator \cite{Lancichinetti2009},  ABCD and ABCD+o (Artificial Benchmark for Community Detection) \cite{abcd,abcdo}, and
nPSO (nonuniform popularity similarity optimization) \cite{Muscoloni2018}.

Given the objective of using synthetic networks to evaluate clustering methods, it is very helpful to have network generators that produce networks with ground truth clusterings that have properties found in clustered real-world networks.
Hence, studies have evaluated synthetic network generators with respect to these properties  \cite{vaca2022systematic,Muscoloni2018}.
An extensive study was reported by  Vaca-Ram{\'\i}rez and Peixoto~\cite{vaca2022systematic}, who evaluated  the graph-tool  \cite{peixoto_graph-tool_2014} software for SBM generation on many real-world networks; their study showed that  SBMs produced using graph-tool had a good fit to many real-world \emph{network properties}, such as degree sequence, local and global clustering coefficients, and diameter.  However, Vaca-Ram{\'\i}rez and Peixoto did not examine whether the ground-truth \emph{clusterings} in the synthetic networks resembled the clusterings of the real-world networks, nor the fit between outlier nodes in the clustered real-world networks and the outlier nodes in the synthetic network.

To motivate this research,  we examined how well SBMs produced using graph-tool are able to reproduce not only the features of a given real-world network but also its clustering.
We find that while these  SBM networks generally have a good fit to the calculated network parameters, they do not exhibit a good fit to edge connectivity within clusters.
Significantly,  the ground truth clusters that are produced for large real-world networks are frequently disconnected.

The importance of having ground truth clusters that are at least connected has already been recognized, and tools have been developed to modify a synthetic network with ground truth clustering to ensure connectedness \cite{viger2016efficient}.
However, to our knowledge,   no methods have yet been developed to address the more ambitious goal of coming close to the edge connectivity of clusters in a given clustered real-world network, while not hurting the fit with respect to other network and clustering properties.
This is the goal of this study.

Here, we present a two-step pipeline for modifying an SBM synthetic network to improve the fit to a given real-world network and clustering, while ensuring also that the network generation still enable some randomness (and so should not just reproduce the given clustered real-world network).
The input to the pipeline is a set of parameters obtained from a clustered real-world network.
In  the first step, we use the REalistic Cluster Connectivity Simulator  ({\sc RECCS}), a new method we present here, which modifies the SBM restricted to the clustered subnetwork in order to improve the fit to the edge connectivity in the real-world clustering.
In the second step, we add in the remaining ``outlier" nodes.
We show that, compared to  the unmodified SBM approach, this two-step approach  produces synthetic networks that have excellent fit for the edge connectivity of the real-world clustered network, while maintaining the fit for other empirical statistics of the clustered real-world network.

\vspace{-.1in}
\section{Materials and Methods}
\label{sec:methods}
\subsection{New Synthetic Network Generation Pipelines}\label{sec:problem}

\paragraph{Software}
The software for {\sc RECCS} and the outlier strategies are available in github at \cite{reccs-github}.

\vspace{-.1in}
\paragraph{High-level description} 

The input is a real-world network $G$ and its clustering. 
Note that the clustering may contain singleton clusters, i.e., clusters containing only a single node; we refer to nodes in singleton clusters as being ``unclustered" or ``outliers", and all other nodes are ``clustered".
The subnetwork of $G$ induced by the non-singleton clusters is called the ``clustered subnetwork" and is denoted by $G_c$.
From this network and clustering, we extract the parameters required for SBM network generation, which includes the assignment of nodes to clusters, number of edges within each cluster and between each pair of clusters, and  the  node degree sequence. We also calculate the edge connectivity of each of the clusters, which we now define.
For a given non-singleton cluster, an edge cut is a set of edges such that deleting those edges, but not the endpoints, disconnects the cluster.
The size of the smallest edge cut for a cluster is its edge connectivity.



We then produce a synthetic network $N$ using  SBM network generation methods in graph-tool \cite{peixoto_graph-tool_2014} to model the clustered subnetwork $G_c$, i.e., if $G$ has any unclustered nodes, then the clustered subnetwork $G_c$ will not be the entire network. Next, we make the SBM network a {\em simple graph} by removing self-loops and replacing each set of parallel edges with a single edge.

We call this modified network $N_c$, noting that it is a reduced version of the original SBM network $N$.
Note that $N_c$  may have many fewer edges than in the real-world network $G_c$, which provides us the opportunity to strategically add edges to ensure the required edge connectivity within the clusters while maintaining the integrity of other network properties; this is Step 1 of the network generation procedure.

If we removed outlier nodes, we also need to add these nodes back into the network, and determine the edges that are incident with these outlier nodes; this is Step 2 of the network generation process, which produces a network only containing edges involving at least one outlier node.
Finally, the two synthetic networks are {\em merged} into one synthetic network.
For additional details, see  the Supplementary Materials \cite{lahari-suppl}.

\vspace{-.15in}
\subsubsection{Step 1: Improving Edge Connectivity.}

We refer to the set of parameters we calculate from the real-world network $G_c$ and its clustering $\mathcal{C}$ as 
 $Param(G_c,\mathcal{C})$.
The problem we seek to solve can be described as follows:
\begin{itemize}
    \item Input: Simple graph $N_c$ with clustering $\mathcal{C}$ and $Param(G_c,\mathcal{C})$, where $\mathcal{C}$ does not have singleton clusters 
    \item Output: Network $N$ with the same clustering $\mathcal{C}$ formed by adding edges to $N_c$,  with the objective of having a good fit to $Param(G_c,\mathcal{C})$.
\end{itemize} 
Recall that the set $Param(G_c \mathcal{C})$ includes the edge connectivity values $k(C)$ for every cluster in $\mathcal{C}$.
Since SBMs in general provide a good fit to many network parameters but not to edge connectivity in clusters, we seek to improve the fit to the $k(C)$ values without hurting the fit to the other parameters.

\begin{figure}[hbt!]%
\centering
\includegraphics[width=0.6\textwidth, height=0.3\textheight]{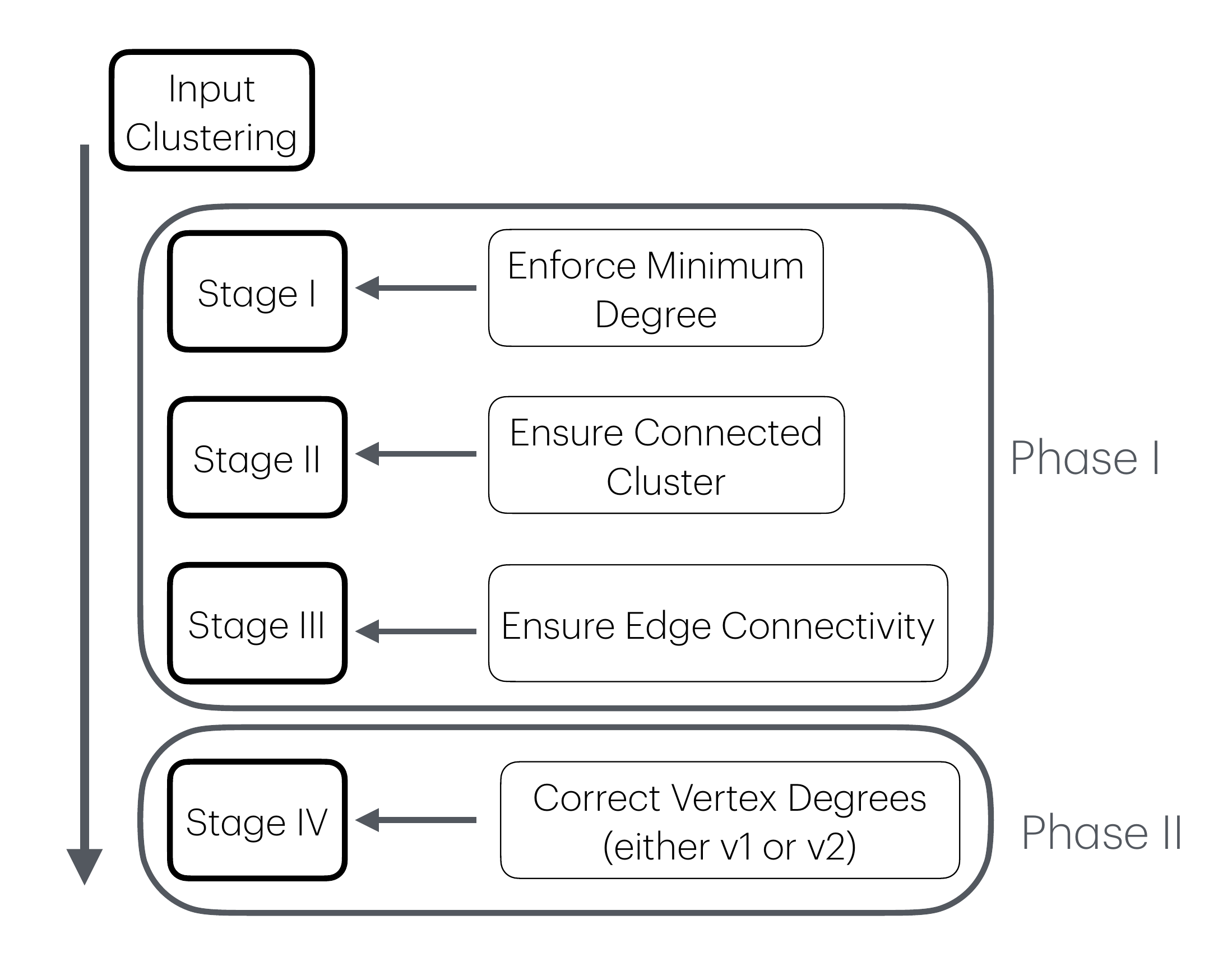}
\caption{\textbf{{\sc RECCS} Workflow.} The two phases of the {\sc RECCS} pipeline, which modifies an initial network (which has no singleton clusters)  by adding edges to improve its fit to the input parameters. The first phase uses the input parameters (clustering, degree sequence, number of edges within and between clusters, and edge connectivity for each cluster) and adds edges within clusters to the starting network to achieve the required edge connectivity, and the second phase adds edges, potentially between clusters, to improve the fit to the degree sequence. See Section \ref{sec:results} for additional details.} 
\label{fig_ce_flow}
\end{figure}

In our experiments, we explored techniques to solve this problem that operate in two basic phases.
In the first phase,  we add edges to ensure that every cluster has edge connectivity $k(C)$, where $\{k(C): C \in \mathcal{C}\}$ is part of part of the input parameter set. 
In the second phase we add additional edges to improve the fit with respect to the degree sequence.
The following two-phase approach, ``{\sc RECCS}", is also outlined in Figure \ref{fig_ce_flow}.

\vspace{-.15in}
\subsubsection{Step 2: Adding in outliers}
 \label{outlierstrategies}



Given the network $G$ and its clustering $\mathcal{C}$, we now focus on creating a synthetic version of the subnetwork $G^*$  of $G$ containing only those edges where at least one endpoint  is an outlier node, i.e., the ``outlier edges", which can connect two outlier nodes or one outlier node and one clustered node. 
Note that $G^*$ includes clustered nodes as well as outlier nodes, and
 has the same set of clusters as in $\mathcal{C}$; however, there are no edges within any non-singleton cluster, and no edges between any two non-singleton clusters.

We propose three strategies for this problem, ranging from Strategy 1, which has the least randomness, to Strategy 3, which has the most randomness.
Each takes the input parameters computed from $G^*$ and $\mathcal{C}$ and then returns a synthetic  network $N^*$ consisting of outlier edges.


\paragraph{Strategy 1:}
This approach has each outlier in its own cluster, and then passes all the parameters computed for $G^*$ to SBM to generate a network. 
Note that this approach reproduces exactly the edges between outlier nodes (as these go between the added singleton clusters) as well as the number of edges between each outlier node and non-singleton cluster. However, there is still randomness in the assignment of edges between outlier nodes and clustered nodes. 

\vspace{-.1in}
\paragraph{Strategy 2: }
All the outliers are placed in a single cluster. 
The set of parameters restricted to the outlier cluster (i.e., the within-cluster degree sequence and number of edges) is used to generate the edges for the outlier cluster.
The remaining edges, between outliers and clustered nodes, are added at random for each outlier node and non-singleton cluster from $\mathcal{C}$ in turn.  
This strategy is more random than Strategy 1 for how it places edges between outlier nodes, but handles edges between outliers and clustered nodes identically as Strategy 1.

\vspace{-.1in}
\paragraph{Strategy 3: }

All the outliers are placed in a single cluster, and the parameters from $G^*$  are used to generate an SBM.

\vspace{-.1in}
\paragraph{Postprocessing: }
We  postprocess $N^*$ as follows: (1) 
if any self-loops or parallel edges have been created, the  excess edges   are removed so that the network becomes a simple graph, and 
(2) if the outlier nodes had been placed in a single cluster during the generation process (i.e., Strategies 2 and 3), then this artificial treatment is ignored, and the outlier nodes are each considered to be singleton clusters.

\vspace{-.1in}
\paragraph{Merging: }
Finally, the random networks generated in Step 1 and Step 2 are merged together; this is straightforward, since the node labelings for each random network are drawn from $G$.

\vspace{-.15in}
\subsection{Datasets}
 The training and test datasets are comprised of clustered networks, with some specifically used for the algorithm design (i.e., training) phase, and others used for the evaluation (i.e., testing) phase.
 
\textbf{Real-world networks}: 
We used \emph{110 real-world networks} of varying sizes and domains. These networks ranged from 1,000 to 13.9 million nodes and are taken from \cite{Netzschleuder} and the SNAP  \cite{leskovec2016snap} repository (see Supplementary Materials \cite{lahari-suppl} for the list of datasets and their properties).   


\textbf{Clustering Methods}: For the network clustering methods, we use Leiden optimizing the Constant Potts Model (CPM) with a range of resolution parameters, Leiden optimizing Modularity, and the Iterative k-core method. Each of these clustering methods is followed by a post-processing technique that removes   clusters below size $11$, then repeatedly breaks up clusters that are not well-connected and reclusters them until all clusters are well-connected (meaning, no cluster $C$ has an edge cut of size at most $\log_{10}(n_C)$, where $n_C$ is the number of nodes in the cluster); effectively this is the same as the Connectivity Modifier without the last stage that removes small clusters \cite{Park2024}. 
 
\textbf{Training Data}: We clustered all 110 networks  using 
Leiden  optimizing the Constant Potts Model (CPM) with resolution parameter $r=0.001$ \cite{Traag2019}, followed by the post-processing described above.
The  clustered networks were used as training data.


\textbf{Testing Data}: We selected  six large real-world networks from the set of 110 networks:  Cit\_hepph, Cit\_patents, the Curated Exosome Network (CEN), Orkut, Wiki\_talk, and Wiki\_topcats, and clustered these networks using  Leiden optimizing CPM with $r=0.01$ and $r=0.1$, Leiden optimizing modularity, and Iterative k-Core (IKC) with $k=10$, each followed by the post-processing described above.
The clustered networks  were used as test data.
Note that because the clusterings are different, 
the testing data are disjoint from the training data. \emph{We are actively generating results from another 104 networks.}


\vspace{-.15in}
\subsection{Evaluation criteria}
We evaluate the similarity between synthetic and real-world networks using various network properties.
The key network properties include the edge connectivity of the clusters (Minimum cuts), network diameter, mixing parameter, degree sequence, 
ratio of disconnected clusters, and clustering coefficients (both global and local). 
We use 
(1) a simple difference, calculated as $(s - s')$, where $s$ is the statistic value of the real-world network and $s'$ is the statistic value of the synthetic network, for scalar properties bounded between 0 and 1, (2) relative difference calculated as ($s$ - $s'$)/$s$ for unbounded scalar statistics, and (3) Root Mean Square Error (RMSE), where RMSE = $\sqrt{\frac{1}{n} \sum_{i=1}^{n} (s_i - s'_i)^2}$, for comparing sequences. 
For  outlier modeling strategies, we examine the outlier degree sequence and the number of edges involving outlier nodes. 
Finally, we report the normalized edit distance between the synthetic network and the real-world network, given the bijection between the node sets.


\vspace{-.15in}
\subsection{Experiments}

\noindent
\textbf{Experiment 1 - Preliminary analysis of SBM:} We compute synthetic networks using SBM given parameters on the training dataset.
The parameters that are given include the node-to-cluster assignment, number of edges within each cluster and between each pair of clusters, and the degree of every node in the network.
The main focus is evaluating   the frequency of disconnected ground truth clusters. 

\textbf{Experiment 2 - Algorithm Design and Development:} 
We explore the algorithm design for the two steps of the synthetic network generation strategy, where  the first step modifies the starting network with respect to  edge connectivity within clusters and  the second step focuses on outlier modeling.
We use the training data for this experiment, and compare our algorithms to SBM generation in graph-tool, post-processed to remove excess edges.

\textbf{Experiment 3 - Evaluation on Test Data:}
In this experiment we evaluate the pipelines we pick in Experiment 2 on testing datasets, in comparison to the generation of SBMs in graph-tool, post-processed to remove excess edges.
 
\begin{figure}[hbt!]%
\centering
\includegraphics[width=0.9\textwidth]{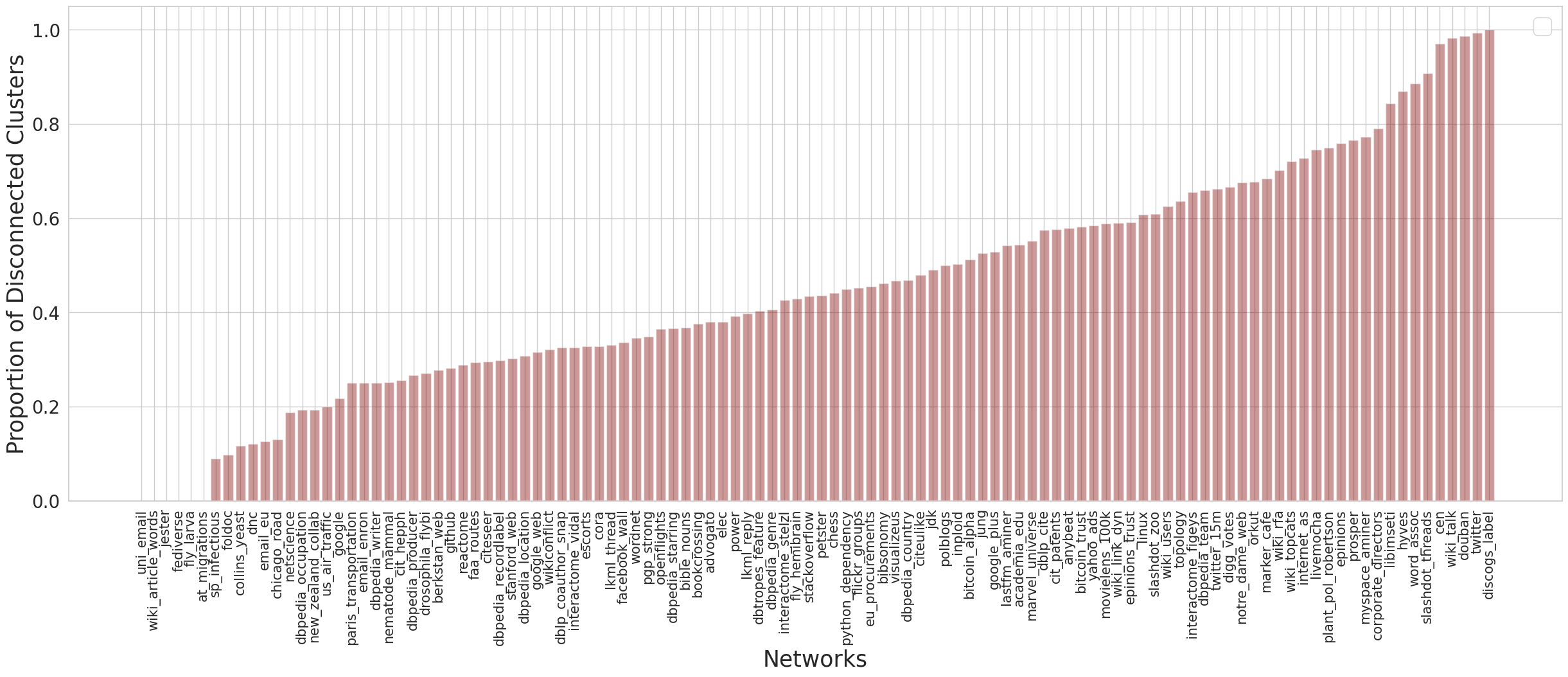}
\caption{\textbf{Proportion of Disconnected clusters in SBM generated networks.}  The x-axis shows 110 SBM networks generated using parameters from real world networks clustered with the Leiden algorithm (training data). Since Leiden clusterings are guaranteed to be connected, this figure shows that SBM method failed to reproduce the connectivity of the real-world clusterings studied here.}\label{fig_disconnect_proportions}
\end{figure}

\vspace{-.1in}
\section{Results and Discussion}
\label{sec:results}
\subsection{Experiment 1 Results: Preliminary Evaluation of SBM}

Figure \ref{fig_disconnect_proportions} shows the proportion of disconnected clusters in the 110 synthetic networks generated by SBM with inputs from training data as detailed in Section \ref{sec:methods}. 
Note that approximately half the networks have more than 40\% of their clusters disconnected. 
Since the input clusterings were based on Leiden, the input clustering parameters given to the SBM generation method in graph-tool are achievable with connected clusters, showing that SBM fails to recover this basic feature of the input clustering.  This trend motivates our study.

\vspace{-.1in}
\subsection{Experiment 2 Results: Design of {\sc RECCS}}\label{exp2} 

The algorithmic structure of {\sc RECCS},  provided in Figure \ref{fig_ce_flow}, is the result of a sequence of experiments on training data  that we now describe. 

{\sc RECCS} takes as input the synthetic network $N_c$ that does not have any singleton clusters and the set $Param(G_c,\mathcal{C})$ of parameter values computed from the clustered subnetwork $G_c$ of a given real-world network $G$ and its clustering $\mathcal{C}$, as described in Section \ref{sec:problem}. 
Here we let $k(C)$ denote the desired edge connectivity for a given cluster $C$.
Because $N_c$ is produced using SBM, there is a ground truth clustering.
Because $N_c$ is formed by deleting edges (in self-loops or extra parallel edges) from an  SBM network, the {\sc RECCS} operates by adding edges, which we do in two phases.
Phase 1 adds edges to ensure that every cluster has at least the desired edge connectivity, and Phase 2 adds edges to improve the fit of the resultant degree sequence.


In our design of {\sc RECCS}  we initially explored techniques that omitted the second phase.
These approaches had excellent fit for the edge connectivity but did not do as well for degree sequence.
We then modified the design of first phase to also consider the degree of the nodes when adding edges to the network.
However, these modifications did not fully address the edge deficit. 
We then introduced the second phase, which adds more edges to improve the degree sequence fit, but noticed that it slightly worsened the mixing parameter fit. 
To counter this effect, we developed a second version of the second phase.
This resulted in  two versions of {\sc RECCS} that differ only in Phase 2, which is when edges are added to improve the fit of the degree sequence.
{\sc RECCS} operates in two phases, as follows.




\begin{itemize}
\item \textbf{Phase 1 (Ensure Edge Connectivity): } Edges are added to each cluster $C$  to ensure edge connectivity at least $k(C)$, as follows:
\begin{itemize}
\item \textbf{Stage 1: Enforce minimum degree:} Here we add edges to ensure that every node has  at least $k(C)$ neighbors in the cluster. 
Therefore, if a node $v$ has $d < k(C)$ neighbors in the cluster, we add $k(C)-d$ edges between $v$ and other nodes in the cluster to which it is not adjacent.

\item \textbf{Stage 2: Ensure connected clusters.} If a cluster $C$ is disconnected, we add $k(C)$ edges at random between its largest component and each of the other components.  

\item \textbf{Stage 3: Ensure edge connectivity. } 
This stage is an iterative method that ensures that the cluster $C$ has edge connectivity at least $k(C)$. Specifically, we use VieCut \cite{henzinger2018practical} to calculate the size of a minimum edge cut; if this size is at least $k(C)$, then we stop (the cluster is sufficiently well connected).
Otherwise, we identify the two parts of the cluster that are connected by fewer than $k(C)$ edges, and we add edges between the two parts.  We then repeat the process until the mincut of the cluster is at least $k(C)$.  

\end{itemize} 
\item \textbf{Phase 2 (Correct vertex degrees): } 
We add edges to
increase the degree of nodes with available degree (i.e., nodes whose current degree is below their target values) using two different techniques:    
\begin{itemize}
\item  $v1$:  for each node with available degree, we add edges to other nodes with available degree, following Algorithm 1 in the Supplementary Materials \cite{lahari-suppl}. 
\item $v2$: edges are strategically added to nodes with available degrees, taking into account the number of inter-cluster and intra-cluster edges in the input subgraph $G_c$. 
This approach, detailed in Algorithm 2 in the Supplementary Materials \cite{lahari-suppl}, restricts the addition of edges to not exceed the expected number of inter-cluster edges. 
\end{itemize}
\end{itemize}

\paragraph{Notes about how we add edges} Recall that we are given a target degree  for every node from $G_c$, but the synthetic network $N$ may not achieve this degree for some nodes.
Those nodes whose current degree is less than the target degree are said to be ``nodes with available degree". 
At each stage of phase $1$ of the algorithm, when adding an edge, we first randomly select nodes within the cluster with available degrees and update their available degree status accordingly. If no suitable nodes with available degree are found, we then randomly choose other nodes within the cluster, even if they have no available degree, to add the edge. 
Note that  we ensure we do not add parallel edges and self-loops.

\begin{figure}[hbt!]%
\centering
\includegraphics[width=0.9\textwidth]{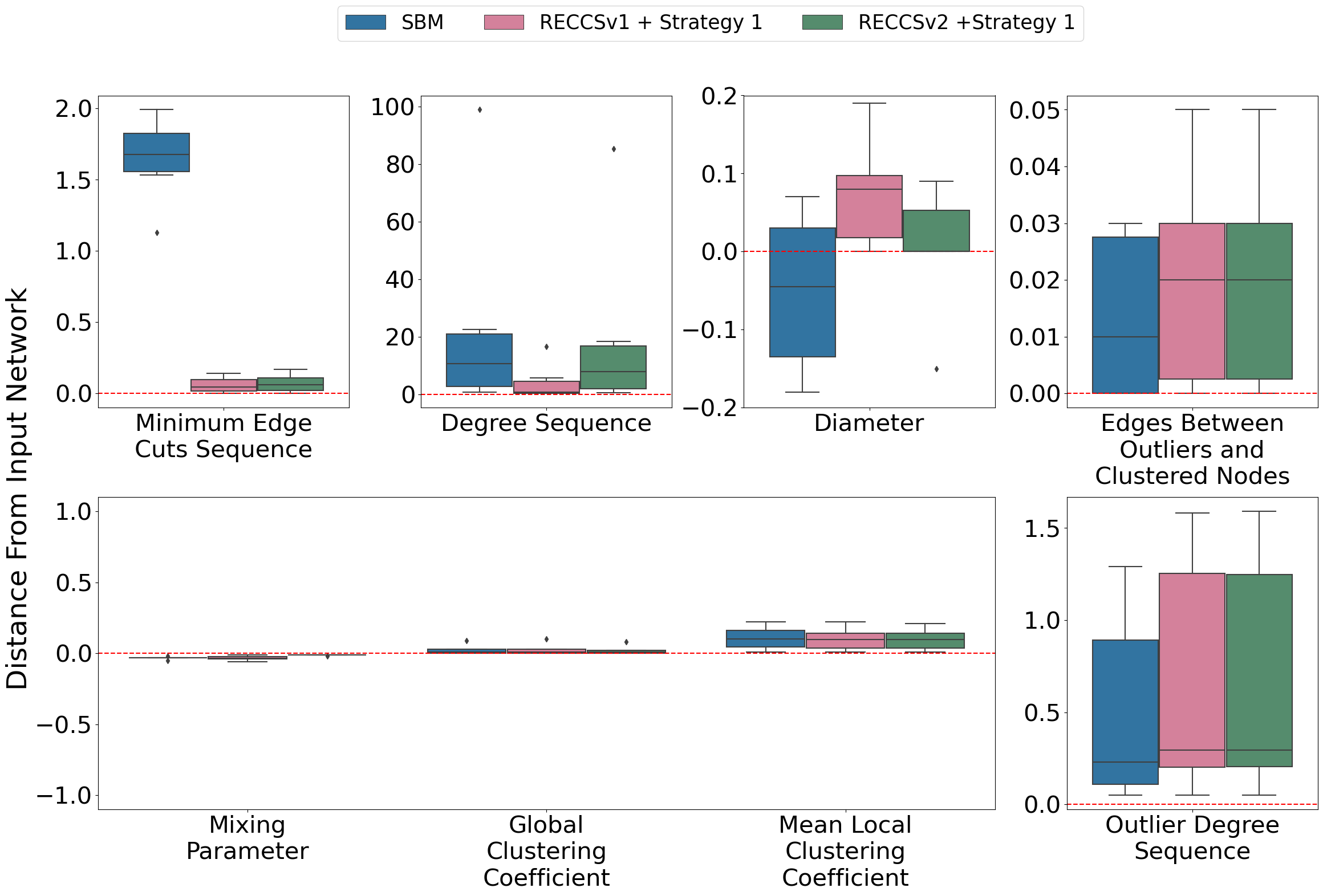}
\caption{\textbf{Comparing SBM to the {\sc RECCS} pipelines on the test networks. }
We compare SBM networks to networks produced using the two pipelines, {\sc RECCS}v1+Strategy 1 and {\sc RECCS}v2+Strategy 1, for different network and clustering statistics. 
The y-axis shows different distance metrics for various network properties. Error is reported using RMSE for degree sequence, outlier degree sequence, and minimum edge cuts sequence; scalar difference is shown for clustering coefficients and mixing parameter; relative difference is shown for the diameter, number of edges between outliers,  and between outliers and clustered nodes. 
The test networks contain six real-world networks, each clustered using
 Leiden-CPM with  $r=0.01$. 
}\label{sbm_ce_os_boxplot}
\end{figure}

\vspace{-.1in}
\subsection{Experiment 3 Results: Evaluation on Test Data}\label{exp3}

Here we show results of the two-step pipelines, each formed by using {\sc RECCS}v1 or {\sc RECCS}v2 for the first step, and then followed by adding in the outlier nodes in three different ways in the second step.
For this experiment, we use the test data only, which has six large real-world networks, each clustered using four different clustering methods (Leiden-CPM with two different values for $r$, Leiden-modularity, and IKC).
Note that the test data are disjoint from the training data.


Since we have two versions for Step 1 and three strategies for Step 2, this produces  six different pipelines.
 A comparison between all six pipelines on the full set of test networks and clusterings is shown in the Supplementary Materials \cite{lahari-suppl}, and reveals that Strategy 1 for Step 2 has the best accuracy. 
Due to space limitations, we present results here just for the two pipelines using outlier Strategy 1.





We begin with the results when clustering using  Leiden-CPM with $r=0.01$ followed by postprocessing, in each case using Strategy 1 for outlier modeling.
As desired, both {\sc RECCS}v1 and {\sc RECCS}v2 substantially improve the fit for the minimum edge cut size compared to SBM (Fig.~\ref{sbm_ce_os_boxplot}, top panel).
We also see an improvement in fit for both versions  for degree sequence compared to SBM (with a larger improvement for {\sc RECCS}v1).
For diameter, {\sc RECCS}v2  improves on the fit compared to SBM, but {\sc RECCS}v1  is slightly worse.
For the other criteria,  the new pipelines have approximately the same accuracy as SBM.
\begin{figure}[hbt!]%
\centering
\includegraphics[width=1\textwidth]{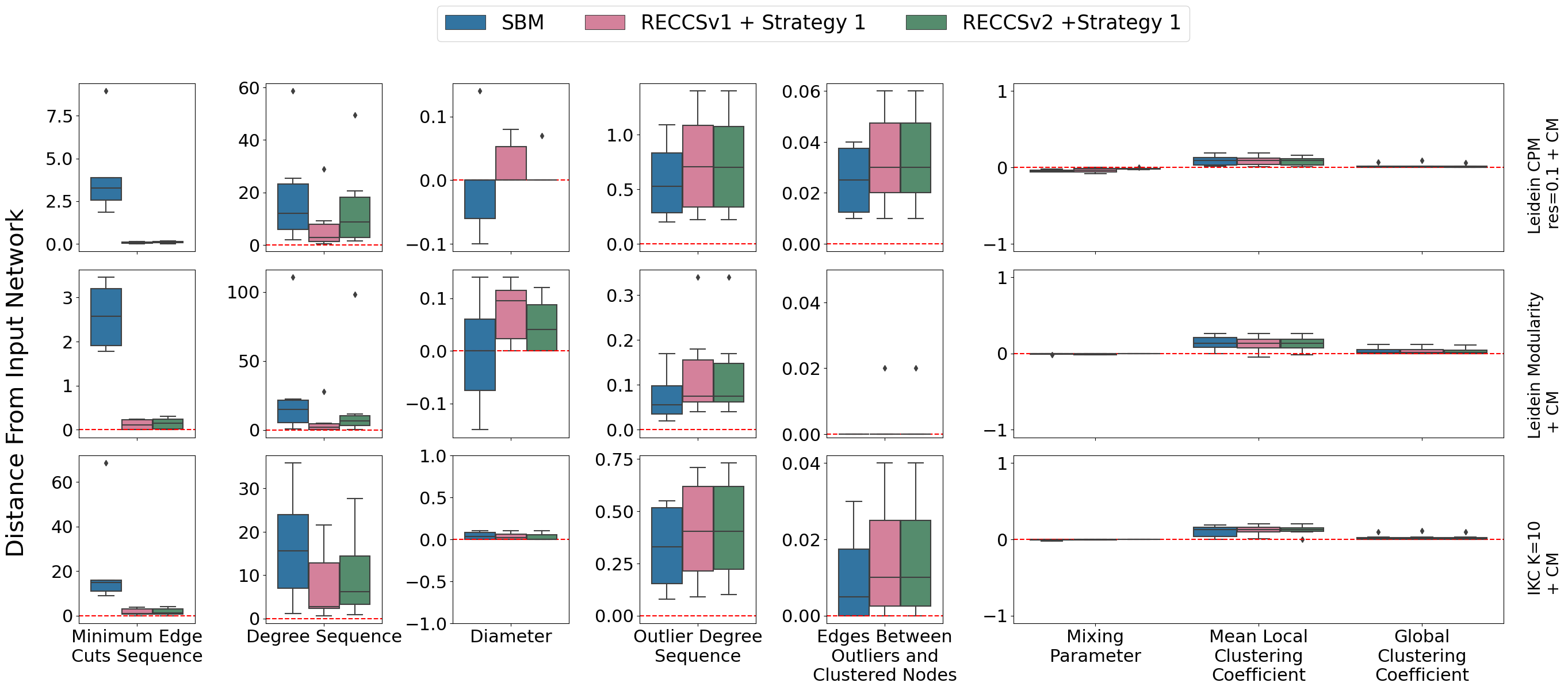}
\caption{\textbf{Accuracy of SBM and Two {\sc RECCS} pipelines on Test Data, using   Three Additional Clusterings} 
The three additional clusterings are Leiden-CPM with $r=0.1$ (top row), Leiden-modularity (middle row), and the Iterative k-core (IKC) method (bottom row).
The y-axis shows different distance metrics for various network properties.  Error is reported using RMSE for degree sequence, outlier degree sequence, and minimum edge cuts sequence; scalar difference is shown for clustering coefficients and mixing parameter; relative difference is shown for the diameter, number of edges between outliers,  and between outliers and clustered nodes. 
}\label{fig:other-test}
\end{figure}

Thus, the two pipelines--{\sc RECCS}v1+Strategy 1 and {\sc RECCS}v2+Strategy 1--both substantially improve on SBM for edge cut sizes, with one clearly better suited for degree sequence and the other clearly better suited for diameter, and were nearly indistinguishable for the other properties.
Results for the other test datasets (Fig.~\ref{fig:other-test}) show the same trends.

\begin{figure}[hbt!]%
\centering
\includegraphics[width=0.9\textwidth]{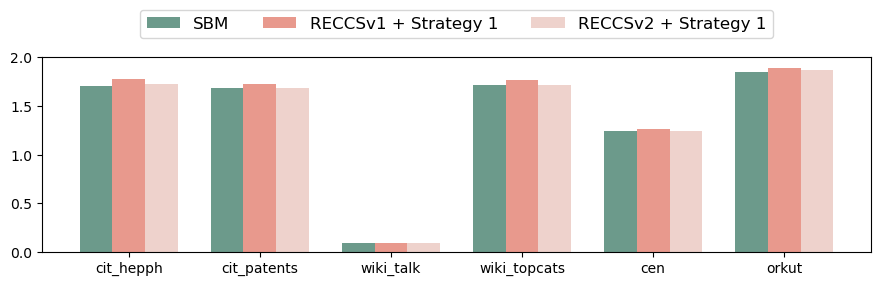}
\caption{\textbf{Comparing SBM, {\sc RECCS}v1,  and {\sc RECCS}v2 with respect to the normalized edit distance between synthetic and real world networks}
The  normalized edit distance between 
the edge sets of the true network $G$ and the synthetic network $N$, i.e., $\frac{|E(G)  \triangle E(N)|} {|E(G)|}$, where $\triangle$ denotes the symmetric difference, and so the maximum possible value is $2.0$. Each real-world network is clustered using
Leiden-CPM, with   $r=0.01$. Here, {\sc RECCS}v2+Strategy 1 produces synthetic networks that are closer to the real-world network that {\sc RECCS}v1+Strategy 1, and about as close as SBM networks. 
}
\label{fig:graph_edit_distance}
\end{figure}
Finally, we compared the two pipelines and unmodified SBMs with respect to the {\em normalized edit distance} between the real-world network and the synthetic network they produce. For this distance, we use the  number of edges that need to be added or removed from the real-world network, to produce the synthetic network, then normalized by the number of edges in the real-world network.
Across the six networks in this test dataset, SBM and {\sc RECCS}v2  are very close, while {\sc RECCS}v2  produced synthetic networks that had a smaller normalized edit distance to the real-world network than {\sc RECCS}v1  (Fig.~\ref{fig:graph_edit_distance}). 
Furthermore, since the maximum normalized edit distance is $2.0$, this shows that in all but one network,  all three synthetic networks have a large enough  distance to the real-world network to clearly not just be close to {\em just replicating} the real-world network.
Thus, all three strategies--unmodified SBM and the two ways of post-processing the SBM--produce networks that are different from the real-world network and have good fit for the network parameters we explored, while the two {\sc RECCS}-pipelines also have a good fit for the cluster edge connectivity values but SBM does not.

\vspace{-.1in}
\section{Conclusion}\label{sec:conclusion}

Motivated by the need for synthetic networks that reproduce features of clustered real-world networks, we introduced the REalistic Cluster Connectivity Simulator ({\sc RECCS}) and  three outlier modeling strategies that can be used to generate synthetic networks.
{\sc RECCS} operates by modifying an SBM synthetic network generated on the clustered nodes in order to improve the fit to the edge-cut sizes for the clusters, and the outlier strategies then add in the remaining unclustered nodes.
We showed, using a diverse set of clustered real-world networks, that the two-step pipelines that use either version of {\sc RECCS} and then outlier Strategy 1  
produce synthetic networks that match or improve on the fit to empirical statistics of the clustered real-world networks compared to SBMs.
Furthermore, the two versions of {\sc RECCS} that we explore have different strengths, allowing for a range of synthetic networks to be developed.

The purpose of this study was to enable the generation of synthetic networks and ground truth clusterings that more closely resemble clustered real-world networks, so that clustering methods can be explored for accuracy on these synthetic networks.
In future work, we will explore a range of clustering methods on these synthetic networks in order to better characterize the conditions under which each method provides accuracy advantages over the other methods.
\vspace{-.15in}
\section*{Funding}
\vspace{-.1in}
This work was supported in part by the  Illinois-Insper Partnership.

\vspace{-.1in}
\bibliographystyle{spmpsci} 
\bibliography{main} 
\end{document}